# MULTIVARIATE INTEGRAL PERTURBATION

# TECHNIQUES - I (THEORY)

**JAN W. DASH**

J. Dash Consultants
jdash9@comcast.net

## Abstract

We present a quasi-analytic perturbation expansion for multivariate $N$-dimensional Gaussian integrals. The perturbation expansion is an infinite series of lower-dimensional integrals (one-dimensional in the simplest approximation). This perturbative idea can also be applied to multivariate Student-t integrals. We evaluate the perturbation expansion explicitly through 2nd order, and discuss the convergence, including enhancement using Padé approximants. Brief comments on potential applications in finance are given, including options, models for credit risk and derivatives, and correlation sensitivities.

## 1. Introduction

The evaluation of multivariate integrals is of substantial interest in various areas of finance including options and credit derivatives, as well as in science and engineering. While some techniques are available in special cases, a common procedure is brute-force numerical integration, e.g. via Monte Carlo simulation.

In this paper, we present a technique that we believe is new and may prove to be useful. It is quasi-analytic and is based on a perturbation expansion. The perturbation expansion gives the $N$-dimensional multivariate Gaussian integral as an infinite series of low-dimensional integrals, which in the simplest case are just one-dimensional integrals. The idea is applicable also to Student-t integrals, and probably others as well. Possibly the expansion, with a clever choice of the initial term, will turn out to provide an interesting and viable numerical approach. This paper discusses theory. Subsequent papers will report on numerical aspects.

The initial "point" about which the expansion is performed is key, as it always is for perturbation expansions. In this work, this "point" is constructed from an approximation to the original correlation matrix involving a factorized expression (one factor), or of sums of such factorized expressions (several factors)[i]. In this paper, we focus on the one-factor approximation. The perturbation expansion is then found in terms of expectation values of {powers of [the difference of (the inverse of the original correlation matrix) and the (inverse of the approximate correlation matrix)]}. These expectation values are with respect to the approximate probability density function, and are readily obtained analytically. We give explicit expressions for Gaussian multivariate integrals of arbitrary dimension using one factor up to second order in the expansion. To 0th order, there is a single one-dimensional integral. In first order, there are $N^2$ additional one-dimensional integrals. In second order, there are $N^4$ additional one-dimensional integrals. All integrals can be grouped into classes of similar appearance, so only a few integrals need to be programmed.

In Sect. 2, we treat the perturbation expansion of multivariate Gaussian integrals for one factor. Sect. 3 has an outline of the procedure for multiple factors. Sect. 4 contains the perturbation expansion of multivariate Student-t integrals. Sect. 5 has some details of the formalism for Gaussian perturbation theory. Sect. 6 deals with the 1st-order perturbation results, and Sect. 7 has the 2nd-order perturbation results. Sect. 8 presents the cluster decomposition diagrammatic notation. Sect. 9 discusses the approximate correlation matrix needed to start the perturbation analysis. Sections 10 and 11 briefly discuss potential applications to options and to credit risk and derivatives. Sect. 12 discusses correlation sensitivity. Sect. 13 discusses convergence, and Sect. 14 discusses enhanced convergence using Padé approximants. Sect. 15 has a logical "flow chart" of the steps needed for numerical investigations.

## 2. Perturbation Expansion of Multivariate Gaussian Integrals

Consider the $N$-dimensional Gaussian multivariate integral $I_N$ with positive-definite correlation matrix $\rho$ with inverse matrix $\rho^{-1}$, determinant $\Delta(\rho)$:

$$I_N\left[\left\{x_i^{\max}\right\};\rho\right]=\left[\Delta(\rho)\right]^{-\frac{1}{2}}\int_{-\infty}^{x_1^{\max}}\cdots\int_{-\infty}^{x_N^{\max}}\exp\left[-\tfrac{1}{2}\boldsymbol{x}^T\rho^{-1}\boldsymbol{x}\right]\,D\boldsymbol{x} \qquad (2.1)$$

Here, $\boldsymbol{x} = \{x_i\}$ is the vector of variables, $\{x_i^{\max}\}$ with $i = 1...N$ are the upper limits assumed constant (i.e. independent of the $\{x_i\}$ variables), and $D\boldsymbol{x} = \prod_{i=1}^{N} \left( dx_i / \sqrt{2\pi} \right)$. Matrix multiplication is understood in the exponent.

Now consider the one-factor approximation $\rho_f$ to the correlation matrix. We introduce numbers $\{c_i\}$ and write for the $i \neq j$ matrix elements of $\rho_f$ :

$$\rho_{f\,ij} = c_i c_j \tag{2.2}$$

It is easy to prove (see Curnow and Dunnett (Ref. i) and Sect. 5) that if we were to replace $\rho$ by $\rho_f$ , then $I_N$ would become a one-dimensional integral

$$I_N\left[\{x_i^{\max}\}; \rho_f\right] = \left[\Delta(\rho_f)\right]^{-\frac{1}{2}} \int_{-\infty}^{x_1^{\max}} \cdots \int_{-\infty}^{x_N^{\max}} \exp\left[-\tfrac{1}{2}\boldsymbol{x}^T \rho_f^{-1} \boldsymbol{x}\right] \; D\boldsymbol{x} \tag{2.3}$$

$$= \int_{-\infty}^{\infty} \frac{d\varsigma}{\sqrt{2\pi}} \exp\left(-\tfrac{1}{2}\varsigma^2\right) \prod_{i=1}^{N} \mathfrak{N}\left[\xi_i^{\max}(\varsigma)\right] \tag{2.4}$$

Here $\varsigma$ is an auxiliary variable. There is one such variable (i.e. one factor), corresponding to the single factorized term $c_i c_j$ for $\rho_{f\,ij}$. Also $\mathfrak{N}(\xi)$ is the standard one-dimensional cumulative normal, and

$$\xi_i^{\max}(\varsigma) = \frac{x_i^{\max} - c_i \varsigma}{s_i} \tag{2.5}$$

Here $s_i = +\sqrt{1 - c_i^2}$ needs to be positive to reproduce $I_N \to 1$ as $\{x_i \to \infty\}$. The single $\varsigma$ integral in Eq. (2.4) can be done straightforwardly using good uniform algebraic approximations for the normal integral $\mathfrak{N}(\xi)$[ii].

The perturbation expansion follows from the trivial identity

$$\rho^{-1} = \rho_f^{-1} + \left(\rho^{-1} - \rho_f^{-1}\right) \tag{2.6}$$

inserted into the original integral. We keep the explicit $\rho_f$ dependence in the exponent as in Eq. (2.3) for $I_N\left[\left\{x_i^{\max}\right\};\rho_f\right]$, and we expand the rest of the exponential in the integrand in a power series in the matrix $\varepsilon$, defined as

$$\varepsilon = \rho^{-1} - \rho_f^{-1} \tag{2.7}$$

So we get

$$I_N\left[\left\{x_i^{\max}\right\};\rho\right] = \sum_{\beta=0}^{\infty} I_N^{(\beta)}\left[\left\{x_i^{\max}\right\};\rho_f;\rho\right] \tag{2.8}$$

where

$$I_N^{(\beta)}\left[\left\{x_i^{\max}\right\};\rho_f;\rho\right] = \frac{1}{\beta\,!}\left(-\tfrac{1}{2}\right)^{\beta} \quad * \\ \frac{1}{\sqrt{\Delta(\rho)}}\int_{-\infty}^{x_1^{\max}} \cdots \int_{-\infty}^{x_N^{\max}} \exp\left[-\tfrac{1}{2}\boldsymbol{x}^T \rho_f^{-1}\boldsymbol{x}\right] \left(\boldsymbol{x}^T \varepsilon \boldsymbol{x}\right)^{\beta} D\boldsymbol{x} \tag{2.9}$$

If we include all the terms in the sum, the result must only depend on the original correlation matrix $\rho$, with all dependence on $\rho_f$ cancelling out.

Using standard functional techniques described in Sect. 5, we can evaluate $I_N^{(\beta)}$ for given $\beta$ as a sum of one-dimensional integrals. For $\beta = 0$ we get the same result up to normalization as the result for $I_N\left[\left\{x_i^{\max}\right\};\rho_f\right]$, namely the single one-dimensional integral

$$I_N^{(\beta=0)}\left[\left\{x_i^{\max}\right\};\rho_f;\rho\right] = J\left(\rho_f,\rho\right)\int_{-\infty}^{\infty}\frac{d\varsigma}{\sqrt{2\pi}}\exp\left(-\tfrac{1}{2}\varsigma^2\right)\prod_{i=1}^{N}\mathfrak{N}\left[\xi_i^{\max}(\varsigma)\right] \tag{2.10}$$

Here the normalization is $J\left(\rho_f,\rho\right) = \sqrt{\Delta\left(\rho_f\right)}\Big/\sqrt{\Delta(\rho)}$ .

For $\beta = 1$ we get one-dimensional integrals of the form

$$I_N^{(\beta=1)}\left[\left\{x_i^{\max}\right\};\rho_f;\rho\right]=-\tfrac{1}{2}J\left(\rho_f,\rho\right)\int_{-\infty}^{\infty}\frac{d\varsigma}{\sqrt{2\pi}}\exp\left(-\tfrac{1}{2}\varsigma^2\right)\ * \qquad (2.11)$$
$$\sum_{i,j=1}^{N}\varepsilon_{ij}G_{ij}\left(\varsigma\right)\prod_{l\neq i,j;\ l=1}^{N}\mathfrak{N}\left[\xi_l^{\max}\left(\varsigma\right)\right]$$

Here, the various $G_{ij}\left(\varsigma\right)$ are functions of $\varsigma$ that are different according to whether $i=j$ or $i\neq j$ (see Sect. 6). The perturbation matrix element is $\varepsilon_{ij}=\left(\rho^{-1}-\rho_f^{-1}\right)_{ij}$. Note that $I_N^{(\beta=1)}$ has an explicit minus sign. There are $N^2$ of these two types of integrals.

For $\beta=2$ we get one-dimensional integrals of the form (see Sect. 7)

$$I_N^{(\beta=2)}\left[\left\{x_i^{\max}\right\};\rho_f;\rho\right]=\tfrac{1}{8}J\left(\rho_f,\rho\right)\int_{-\infty}^{\infty}\frac{d\varsigma}{\sqrt{2\pi}}\exp\left(-\tfrac{1}{2}\varsigma^2\right)\ * \qquad (2.12)$$
$$\sum_{i,j,k,l=1}^{N}\varepsilon_{ij}\varepsilon_{kl}G^{(\Gamma)}_{(ijkl)\text{ in }C(\Gamma)}\left(\varsigma\right)$$

Here $\Gamma$ is a "class label" corresponding to the seven different classes $C\left(\Gamma\right)$ of indices $\left(i,j,k,l\right)$ that arise when the indices are all distinct, have one pair equal, two pairs equal, three values equal, or four values equal. The classes can be conveniently associated with diagrams, in a sort of "cluster decomposition" [iii, 1]. The various functions $G^{(\Gamma)}_{(ijkl)\text{ in }C(\Gamma)}$ are categorized using the class label $\Gamma$. There are a total of $N^4$ of these integrals, of five different types. Because the indexing is complicated, products of normal functions are lumped into $G^{(\Gamma)}_{(ijkl)\text{ in }C(\Gamma)}$.

## 3. Multiple Factors for the Gaussian Perturbation Expansion

Using multiple-factors means we take $\rho$ approximated by $\rho_f$ of the form :

$$\rho_{fij}=\sum_{\alpha=1}^{K}c_{i\alpha}c_{j\alpha}\ \text{ for }\left(i\neq j\right) \qquad (3.1)$$

[1] **Cluster Decomposition:** The cluster decomposition in the present paper is similar in spirit but different in detail from the cluster decomposition for multivariate Gaussian integrals that I envisioned in the late 1980's (see p. 623 of Ref. iii).

with $\alpha = 1...K$. This leads to $K$-dimensional integrals using $K$ auxiliary variables $\{\varsigma_\alpha\}$ that replace the one-dimensional $\varsigma$ integrals. Perturbation theory is performed along the same lines. The tradeoff is between the extra complexity of multidimensional integrals versus a better initial description of the correlation matrix. If $K = N$ and with principal components to get $\rho_f$ (as explained in Sect. 9), we restore the original problem, and nothing is accomplished.

We shall, however, only give results for perturbations about a one-factor approximation in this paper.

## 4. Perturbation Expansion of Multivariate Student-t Integrals

We next outline perturbation theory for multivariate Student-t integrals. This produces two-dimensional integrals instead of one-dimensional integrals, in the one-factor approximation to $\rho_f$. The basic idea is to use an integral representation in order to rewrite things in exponential form, so that we can again use perturbation theory similarly to the Gaussian case.

The multivariate Student-t ("St") cumulative probability distribution in $N$ dimensions with $\nu$ degrees of freedom is[iv]

$$I_N^{(St)}\left[\left\{x_i^{\max}\right\};\rho;\nu\right] = \frac{K^{(1)}}{\left[\Delta(\rho)\right]^{\frac{1}{2}}} \int_{-\infty}^{x_1^{\max}} \cdots \int_{-\infty}^{x_N^{\max}} \left(1+\tfrac{1}{\nu}\boldsymbol{x}^T\rho^{-1}\boldsymbol{x}\right)^{-(\nu+N)/2} D\boldsymbol{x} \qquad (4.1)$$

with $K^{(1)} = \Gamma\left[\tfrac{1}{2}(\nu+N)\right]\left(\tfrac{1}{2}\nu\right)^{-\frac{N}{2}}\left\{\Gamma\left[\tfrac{1}{2}\nu\right]\right\}^{-1}$. We use the identity

$$\left(1+\tfrac{1}{\nu}\boldsymbol{x}^T\rho^{-1}\boldsymbol{x}\right)^{-(\nu+N)/2} = K^{(2)}\int_0^\infty y^{\nu+N-1}\exp\left[-\tfrac{1}{2}\left(1+\tfrac{1}{\nu}\boldsymbol{x}^T\rho^{-1}\boldsymbol{x}\right)y^2\right]dy \qquad (4.2)$$

with $K^{(2)} = \left\{\Gamma\left[\tfrac{1}{2}(\nu+N)\right]2^{(\nu+N-2)/2}\right\}^{-1}$. The integral over $y$ is an extra complication for the Student-t distribution. We make the change of variables $u_i = yx_i/\sqrt{\nu}$ and define $u_i^{\max}(y) = yx_i^{\max}/\sqrt{\nu}$. We obtain

$$I_N^{(St)}\left[\left\{x_i^{\max}\right\};\rho;\nu\right]=\frac{2^{1-\frac{1}{2}\nu}}{\left[\Delta(\rho)\right]^{\frac{1}{2}}\Gamma\left[\frac{1}{2}\nu\right]}\int_0^\infty y^{\nu-1}e^{-\frac{1}{2}y^2} * \left[\int_{-\infty}^{u_1^{\max}(y)}\cdots\int_{-\infty}^{u_N^{\max}(y)}\exp\left[-\tfrac{1}{2}\boldsymbol{u}^T\rho^{-1}\boldsymbol{u}\right]D\boldsymbol{u}\right]dy \tag{4.3}$$

Here $D\boldsymbol{u}=\prod_{i=1}^{N}\left(du_i/\sqrt{2\pi}\right)$. We now have a Gaussian form and thus can use the same perturbation procedure in Sect. I. We expand in powers of $\varepsilon=\rho^{-1}-\rho_f^{-1}$ and perform the integrals over the $\boldsymbol{u}$ variables with respect to the measure with $\exp\left[-\frac{1}{2}\boldsymbol{u}^T\rho_f^{-1}\boldsymbol{u}\right]$. Assuming a one factor approximation for $\rho_f$, this again yields one-dimensional integrals over $\varsigma$, which along with the integral over $y$, gives the perturbative result for $I_N^{(St)}\left[\left\{x_i^{\max}\right\};\rho;\nu\right]$ in terms of sums of two-dimensional integrals. Explicitly,

$$I_N^{(St)}\left[\left\{x_i^{\max}\right\};\rho;\nu\right]=\frac{2^{1-\frac{1}{2}\nu}}{\Gamma\left[\frac{1}{2}\nu\right]}\int_0^\infty y^{\nu-1}e^{-\frac{1}{2}y^2}I_N\left[\left\{\frac{yx_i^{\max}}{\sqrt{\nu}}\right\};\rho\right]dy \tag{4.4}$$

Here $I_N\left[\left\{\frac{yx_i^{\max}}{\sqrt{\nu}}\right\};\rho\right]$ is the Gaussian multivariate Eq. (2.1) with changed upper limits for fixed $y$, namely $x_i^{\max}\to yx_i^{\max}/\sqrt{\nu}=u_i^{\max}(y)$. We have

$$I_N\left[\left\{u_i^{\max}(y)\right\};\rho\right]=\left[\Delta(\rho)\right]^{-\frac{1}{2}}\int_{-\infty}^{u_1^{\max}(y)}\cdots\int_{-\infty}^{u_N^{\max}(y)}\exp\left[-\tfrac{1}{2}\boldsymbol{u}^T\rho^{-1}\boldsymbol{u}\right]D\boldsymbol{u} \tag{4.5}$$

We can now use the results from the previous analysis for multivariate Gaussian perturbation theory, inserted into Eq. (4.4), to get the perturbation expansion for the multivariate Student-t integral, including the $y$ integral.

For large $\nu$, it is well known that the Student-t approaches the Gaussian. Using WKB around $y^*=\sqrt{\nu}$ eliminates the $y$ integral, and yields the result:

$$I_N^{(St)}\left[\left\{x_i^{\max}\right\};\rho;\nu\right]\underset{\text{Large }\nu}{\overset{WKB}{\approx}}I_N\left[\left\{x_i^{\max}\right\};\rho\right] \tag{4.6}$$

## 5. Formalism for Gaussian Perturbation Theory

This section discusses some details of the formalism for Gaussian perturbation theory. We first show how the extra variable $\varsigma$ arises. Start with Eq. (2.3) and note the identity

$$\exp\left[-\tfrac{1}{2}\boldsymbol{x}^T \rho_f^{-1}\boldsymbol{x}\right] = \sqrt{\Sigma^2}\int_{-\infty}^{\infty}\frac{d\varsigma}{\sqrt{2\pi}}\exp\left(-\tfrac{1}{2}\varsigma^2\right)\prod_{i=1}^{N}\exp\left[-\tfrac{1}{2}\frac{\left(x_i - c_i\varsigma\right)^2}{s_i^2}\right] \quad (5.1)$$

Here

$$\Sigma^2 = 1 + \sum_{l=1}^{N}\frac{c_l^2}{s_l^2} \quad (5.2)$$

This identity introduces the extra variable $\varsigma$. Now we use Eq. (5.1) in Eq.(2.3), and change variables, introducing $\{\xi_i\}$ to eliminate the $\{x_i\}$:

$$\xi_i = \frac{x_i - c_i\varsigma}{s_i} \text{ or } x_i = c_i\varsigma + s_i\xi_i \quad (5.3)$$

The integral in Eq. (2.3) becomes

$$I_N\left[\{x_i^{\max}\};\rho_f\right] = \int_{-\infty}^{\infty}\frac{d\varsigma}{\sqrt{2\pi}}\exp\left(-\tfrac{1}{2}\varsigma^2\right)\prod_{i=1}^{N}\left\{\int_{-\infty}^{\xi_i^{\max}(\varsigma)}\frac{d\xi_i}{\sqrt{2\pi}}\exp\left[-\tfrac{1}{2}\xi_i^2\right]\right\} \quad (5.4)$$

where $\xi_i^{\max}(\varsigma)$ is in Eq. (2.5). This immediately yields Eq. (2.4).

Consider the average $\langle F\rangle_{\rho_f}$ of any function $F\left[\{x_i\}\right]$ with respect to the approximation measure with $\rho_f$ in Eq. (2.3). This is:

$$\langle F\rangle_{\rho_f} = \left[\Delta\left(\rho_f\right)\right]^{-\frac{1}{2}}\int_{-\infty}^{x_1^{\max}}\cdots\int_{-\infty}^{x_N^{\max}} F\left[\{x_i\}\right]\exp\left[-\tfrac{1}{2}\boldsymbol{x}^T\rho_f^{-1}\boldsymbol{x}\right]D\boldsymbol{x} \quad (5.5)$$

Such integrals can be handled using a standard functional technique. We introduce “currents” $\{J_i\}$ and functional derivatives $\{\partial/\partial J_i\}$. Making the same change of variables in Eq. (5.3), we get

$$\left\langle F \right\rangle_{\rho_f} = \int_{-\infty}^{\infty} \frac{d\varsigma}{\sqrt{2\pi}} \exp\left(-\tfrac{1}{2}\varsigma^2\right) F\left[\left\{ c_i\varsigma + s_i \frac{\partial}{\partial J_i} \right\}\right] \; * \qquad (5.6)$$
$$\prod_{i=1}^{N} \left\{ \int_{-\infty}^{\xi_i^{\max}(\varsigma)} \frac{d\xi_i}{\sqrt{2\pi}} \exp\left[-\tfrac{1}{2}\xi_i^2 + \xi_i J_i\right] \right\} \Big|_{\{J_i = 0\}}$$

Restoring the original $\left\{ x_i^{\max} \right\}$ upper limit parameters, we get

$$\left\langle F \right\rangle_{\rho_f} = \int_{-\infty}^{\infty} \frac{d\varsigma}{\sqrt{2\pi}} \exp\left(-\tfrac{1}{2}\varsigma^2\right) F\left[\left\{ c_i\varsigma + s_i \frac{\partial}{\partial J_i} \right\}\right] \; * \qquad (5.7)$$
$$\prod_{i=1}^{N} \left\{ \exp\left(\tfrac{1}{2} J_i^2\right) \mathfrak{N}\left[\left(\frac{x_i^{\max} - c_i\varsigma}{s_i}\right) - J_i\right] \right\} \Big|_{\{J_i = 0\}}$$

For the problem at hand, we set

$$F\left[\left\{ x_i \right\}\right] = \sum_{\beta=0}^{\infty} \frac{1}{\beta\,!} \left(-\tfrac{1}{2}\right)^{\beta} \left(\boldsymbol{x}^T \varepsilon \boldsymbol{x}\right)^{\beta} \qquad (5.8)$$

Writing out the components of the matrices, we have

$$\left(\boldsymbol{x}^T \varepsilon \boldsymbol{x}\right)^{\beta} = \sum_{\substack{\{i_\gamma, j_\gamma\}=1 \\ \gamma=1\ldots\beta}}^{N} \left[ \prod_{\gamma=1}^{\beta} x_{i_\gamma} \varepsilon_{i_\gamma j_\gamma} x_{j_\gamma} \right] \qquad (5.9)$$

We make the replacement $x_i \to c_i\varsigma + s_i\, \partial/\partial J_i$, evaluate the $\left\{ J_i \right\}$ derivatives and then set $\left\{ J_i = 0 \right\}$. Sects. 6 and 7 contain the results for $\beta \le 2$.

It is useful to have an analytic inverse for the one-factor matrix $\rho_f$. This is[v]

$$\left(\rho_f\right)_{ii}^{-1} = \frac{1}{s_i^2}\left(1 - \frac{c_i^2}{s_i^2 \Sigma^2}\right) \qquad (5.10)$$

$$\left(\rho_f\right)_{ij}^{-1} = -\frac{c_i c_j}{s_i^2 s_j^2 \Sigma^2} \text{ for } \left(i \ne j\right) \qquad (5.11)$$

where $\Sigma^2$ is given in Eq. (5.2). For multiple factors, the inverse of $\rho_f$ is complicated, and we have been unable to find an analytic inverse.

## 6. Details of the $\beta = 1$ (1st order) terms

In this section, we give the $\beta = 1$ functions $G_{ij}(\varsigma)$ in the text. First, define

$$\nu_i(\varsigma) = \mathfrak{N}\left[\xi_i^{\max}(\varsigma)\right] \tag{6.1}$$

and

$$\chi_i(\varsigma) = -\frac{1}{\sqrt{2\pi}} \exp\left\{\tfrac{1}{2}\left[\xi_i^{\max}(\varsigma)\right]^2\right\} \tag{6.2}$$

Also define

$$w_i^{(1)}(\varsigma) = c_i \varsigma \nu_i(\varsigma) + s_i \chi_i(\varsigma) \tag{6.3}$$

and

$$w_i^{(2)}(\varsigma) = \left(c_i^2 \varsigma^2 + s_i^2\right)\nu_i(\varsigma) + \left[2 c_i s_i \varsigma + s_i^2 \xi_i^{\max}(\varsigma)\right]\chi_i(\varsigma) \tag{6.4}$$

Then for equal indices $i = j$

$$G_{ii}(\varsigma) = w_i^{(2)}(\varsigma) \tag{6.5}$$

and for unequal indices $i \neq j$

$$G_{ij}(\varsigma)\,\Big|_{i\neq j} = w_i^{(1)}(\varsigma)\, w_j^{(1)}(\varsigma) \tag{6.6}$$

## 7. Details of the $\beta = 2$ (2nd order) terms

In this Section, we discuss the 2nd-order $\beta = 2$ terms. We first define $w_i^{(3)}(\varsigma)$ and $w_i^{(4)}(\varsigma)$:

$$w_i^{(3)}(\varsigma) = \left(c_i^3\varsigma^3 + 3c_i s_i^2 \varsigma\right)\nu_i(\varsigma) + \left\{3c_i^2 s_i \varsigma^2 + 3c_i s_i^2 \varsigma \xi_i^{\max}(\varsigma) + s_i^3\left[\left(\xi_i^{\max}(\varsigma)\right)^2 + 2\right]\right\}\chi_i(\varsigma) \tag{7.1}$$

$$w_i^{(4)}(\varsigma) = \left(c_i^4\varsigma^4 + 6c_i^2 s_i^2 \varsigma^2 + 3s_i^4\right)\nu_i(\varsigma) + \left\{\begin{array}{l} 4c_i^3 s_i \varsigma^3 + 6c_i^2 s_i^2 \varsigma^2 \xi_i^{\max}(\varsigma) + 4c_i s_i^3 \varsigma\left[\left(\xi_i^{\max}(\varsigma)\right)^2 + 2\right] \\ + s_i^4 \xi_i^{\max}(\varsigma)\left[\left(\xi_i^{\max}(\varsigma)\right)^2 + 3\right] \end{array}\right\}\chi_i(\varsigma) \tag{7.2}$$

The seven classes of terms are defined as follows corresponding to the indices $(i, j, k, l)$ in the outer product of the two $\varepsilon$ matrices with elements $\varepsilon_{ij}\varepsilon_{kl}$. Two of them are degenerate, so there are really only five independent classes.

Class $\Gamma = 1$: No two indices are equal. There are $N(N-1)(N-2)(N-3)$ such terms for $N \geq 4$. An example is $\varepsilon_{12}\varepsilon_{34}$. The corresponding function is $G_{1234}^{(\Gamma=1)}(\varsigma) = w_1^{(1)}(\varsigma) w_2^{(1)}(\varsigma) w_3^{(1)}(\varsigma) w_4^{(1)}(\varsigma) \prod_{m \neq 1,2,3,4}^{N} \nu_m(\varsigma)$. In general,

$$G_{ijkl}^{(\Gamma=1)}(\varsigma) = w_i^{(1)}(\varsigma) w_j^{(1)}(\varsigma) w_k^{(1)}(\varsigma) w_l^{(1)}(\varsigma) \prod_{m \neq i,j,k,l}^{N} \nu_m(\varsigma) \tag{7.3}$$

Class $\Gamma = 2$: One pair of indices is equal in one $\varepsilon$ matrix; the others are unequal to each other and unequal to the pair. There are $2N(N-1)(N-2)$ such terms for $N \geq 3$. An example is $\varepsilon_{11}\varepsilon_{23}$. The corresponding function is $G_{1123}^{(\Gamma=2)}(\varsigma) = w_1^{(2)}(\varsigma) w_2^{(1)}(\varsigma) w_3^{(1)}(\varsigma) \prod_{m \neq 1,2,3}^{N} \nu_m(\varsigma)$. In general,

$$G_{iijk}^{(\Gamma=2)}(\varsigma) = w_i^{(2)}(\varsigma) w_j^{(1)}(\varsigma) w_k^{(1)}(\varsigma) \prod_{m \neq i,j,k}^{N} \nu_m(\varsigma) \tag{7.4}$$

Class $\Gamma = 3$: One pair of indices is crosswise equal between the two $\varepsilon$ matrices; the others are unequal to each other and unequal to the pair. There are $4N(N-1)(N-2)$ such terms for $N \geq 3$. An example is $\varepsilon_{12}\varepsilon_{13}$. The corresponding function is degenerate, $G_{1213}^{(\Gamma=3)}(\varsigma) = G_{1123}^{(\Gamma=2)}(\varsigma)$. In general,

$$G_{ijik}^{(\Gamma=3)}(\varsigma) = G_{iijk}^{(\Gamma=2)}(\varsigma) \qquad (7.5)$$

Class $\Gamma = 4$: Each pair of indices on each $\varepsilon$ matrix are equal, but distinct. There are $N(N-1)$ such terms for $N \geq 2$. An examples is $\varepsilon_{11}\varepsilon_{22}$. The corresponding function is

$G_{1122}^{(\Gamma=4)}(\varsigma) = w_1^{(2)}(\varsigma) w_2^{(2)}(\varsigma) \prod_{m \neq 1,2}^{N} \nu_m(\varsigma)$, or in general

$$G_{iijj}^{(\Gamma=4)}(\varsigma) = w_i^{(2)}(\varsigma) w_j^{(2)}(\varsigma) \prod_{m \neq i,j}^{N} \nu_m(\varsigma) \qquad (7.6)$$

Class $\Gamma = 5$: Two pairs of indices are crosswise equal, but distinct. There are $2N(N-1)$ such terms for $N \geq 2$. An example is $\varepsilon_{12}\varepsilon_{12}$. The corresponding function is degenerate, $G_{1212}^{(\Gamma=5)}(\varsigma) = G_{1122}^{(\Gamma=4)}(\varsigma)$. In general,

$$G_{ijij}^{(\Gamma=5)}(\varsigma) = G_{iijj}^{(\Gamma=4)}(\varsigma) \qquad (7.7)$$

Class $\Gamma = 6$: Three indices are equal, different from the fourth. An example is $\varepsilon_{11}\varepsilon_{12}$. There are $4N(N-1)$ such terms for $N \geq 2$. An example is $\varepsilon_{11}\varepsilon_{12}$. The corresponding function is $G_{1112}^{(\Gamma=6)}(\varsigma) = w_1^{(3)}(\varsigma) w_2^{(1)}(\varsigma) \prod_{m \neq 1,2}^{N} \nu_m(\varsigma)$. In general,

$$G_{iiij}^{(\Gamma=6)}(\varsigma) = w_i^{(3)}(\varsigma) w_j^{(1)}(\varsigma) \prod_{m \neq i,j}^{N} \nu_m(\varsigma) \qquad (7.8)$$

Class $\Gamma = 7$: All indices are equal. There are $N$ terms. An example is $\varepsilon_{11}\varepsilon_{11}$. The corresponding function is $G_{1111}^{(\Gamma=7)}(\varsigma) = w_1^{(4)}(\varsigma) \prod_{m \neq 1}^{N} \nu_m(\varsigma)$. In general

$$G_{iiii}^{(\Gamma=7)}(\varsigma) = w_i^{(4)}(\varsigma) \prod_{m \neq i}^{N} \nu_m(\varsigma) \qquad (7.9)$$

The total number of terms from all classes is $N^4$, and as is clear from the above discussion there are five types of integrals with different parameters.

## 8. Cluster Decomposition Diagrammatic Notation

The cluster decomposition is a useful device that is a visual mnemonic for the various types of terms (cf. Ref. iii, p.700). Each index is represented by a line going from left to right. Each pair of lines starting at the left and going from top to bottom corresponds to indices $i, j$ on one of the $\varepsilon_{ij}$ matrices. Consider Fig. 1 below. If two indices are equal, we make the lines enter and leave a “bubble”, like the picture on the left. If two indices are unequal, they will not go into the same “bubble”, like the picture on the right.

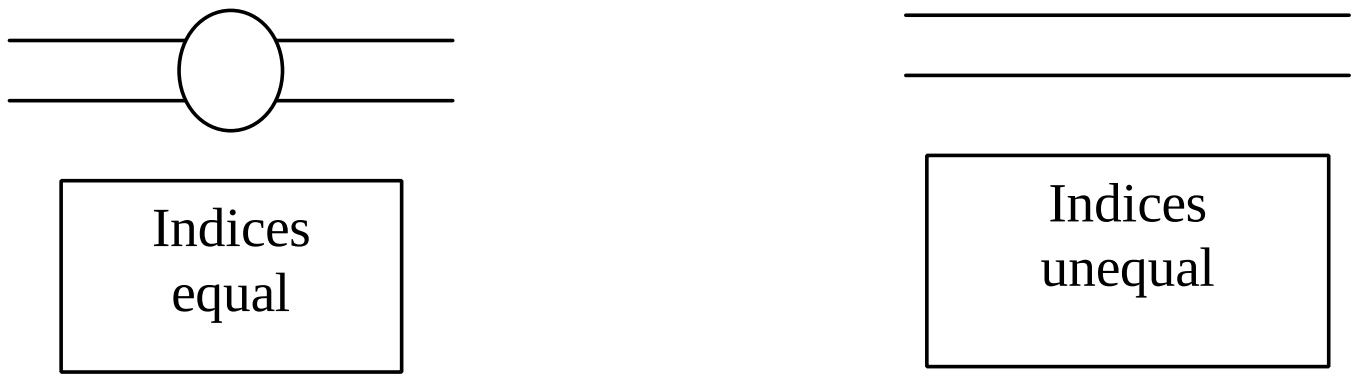

**Fig. 1: 1st-order diagrams**

The seven classes in the 2nd order perturbation term with two $\varepsilon_{ij}\varepsilon_{kl}$ matrices with four indices (so, four lines) can be drawn as in Fig. 2 below.

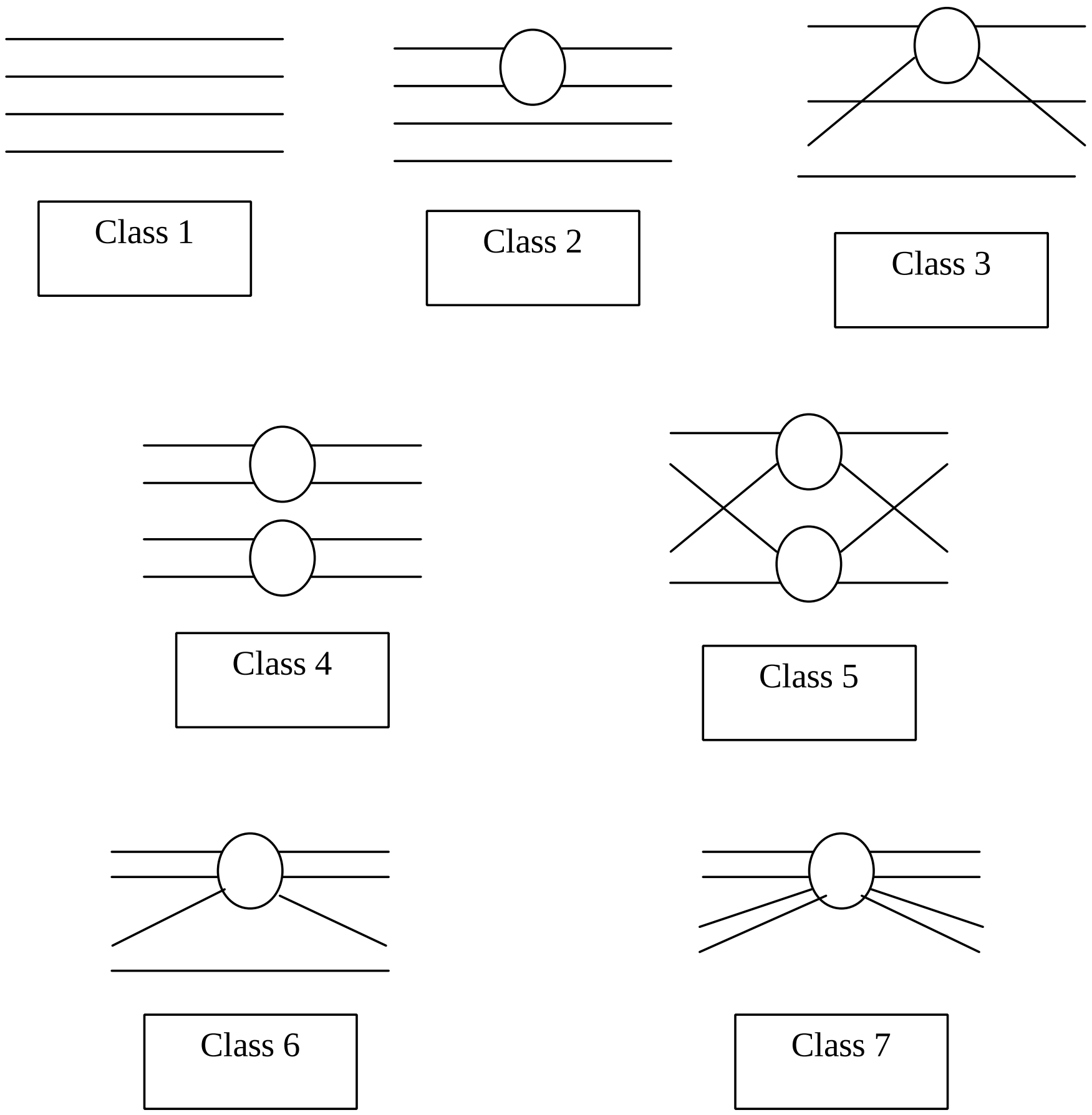

**Fig. 2: 2nd-order diagrams**

## 9. Choice of the parameters in the matrix $\rho_f$

The choice of the $\{c_{i\alpha}\}$ parameters in the $K$ -factor approximation matrix $\rho_f$ is arbitrary. Of course the perturbation results to some given order will depend on the choice. One idea is to use a principal component (PC) decomposition of the original correlation matrix $\rho$, keep the first $K$ of the PCs, and thereby identify $\{c_{i\alpha}\}$. The PC decomposition of the $NxN$ matrix $\rho$ is a sum of $N$ factorized terms, and is an identity. It reads

$$\rho_{ij} = \sum_{\mu=1}^{N} \psi_i^{(\mu)} \lambda^{(\mu)} \psi_j^{(\mu)} \tag{9.1}$$

Here $\psi^{(\mu)}$ and $\lambda^{(\mu)}$ are the $\mu$ -th eigenfunction and eigenvalue of $\rho$ . This is already a sum of factorized terms in the matrix indices. Hence, truncating the sum at the $K$ th term yields a $K$ - factor approximation $\rho_f$ to $\rho$ . This is exactly the procedure used in VAR calculations using SVD to get rid of $N-K$ negative eigenvalues, setting them to zero. We need to renormalize the eigenfunctions so that $\rho_f$ will have unit diagonal elements (cf. p. 327 of Ref. iii). So we replace $\psi_i^{(\mu)}$ by $\tilde{\psi}_i^{(\mu)} = \gamma_i \psi_i^{(\mu)}$, where $\gamma_i = \left[ \sum_{\mu=1}^{K} \left( \psi_i^{(\mu)} \right)^2 \lambda^{(\mu)} \right]^{-\frac{1}{2}}$ , and then write

$$\rho_{fij} = \sum_{\alpha=1}^{K} c_{i\alpha} c_{j\alpha} = \sum_{\mu=1}^{K} \tilde{\psi}_i^{(\mu)} \lambda^{(\mu)} \tilde{\psi}_j^{(\mu)} \tag{9.2}$$

Identifying terms with $\mu = \alpha$ , we then can set $c_{i\alpha} = \tilde{\psi}_i^{(\alpha)} \sqrt{\lambda^{(\alpha)}}$ . If $\rho$ is positive definite with $N$ positive eigenvalues $\left\{ \lambda^{(\mu)} > 0 \right\}$, then the $NxN$ matrix $\rho_f$ is positive semi-definite with $K$ positive eigenvalues and $N-K$ zero eigenvalues.

For one factor, $K = 1$, this procedure just produces $\rho_{fij} = 1$, which is not too useful. However, we could really assume any constant $\rho_{fij} > -1/(N-1)$ for $i \neq j$ , maintaining $\rho_f$ positive definiteness. To do this, we can look for a best fit to $\rho$ for the constant approximation matrix $\rho_f$ with $\rho_{fij} = c^2$ for $i \neq j$ . Hence, write $\chi^2 = \sum_{i \neq j=1}^{N} \left( c^2 - \rho_{ij} \right)^2$ and set $\partial \chi^2 / \partial c^2 = 0$ . We get

$$c = \left[ \frac{1}{N(N-1)} abs \left( \sum_{i,l \neq i} \rho_{il} \right) \right]^{\frac{1}{2}} \tag{9.3}$$

Here, " $abs$ " means absolute value; we include this so that the expression always makes sense.

A generalization to a nonconstant matrix $\rho_f$ can be done by pulling out the sum over the index $i$ in Eq. (9.3), and then identifying $c_i$ for given $i$ as

$$c_i = \left[ \frac{1}{(N-1)} abs\left( \sum_{l \neq i} \rho_{il} \right) \right]^{\frac{1}{2}} \operatorname{sgn}\left[ \sum_{l \neq i} \rho_{il} \right] \tag{9.4}$$

Here, "$\operatorname{sgn}$" means sign, and is included for generality so that $\rho_f$ can have matrix elements with signs. While this procedure may seem arbitrary, again recall that we are free to choose the approximate correlation matrix $\rho_f$ any way we want. With Eq. (9.4), we get the final result for the one factor approximate matrix elements $(i \neq j)$ that we propose to use,

$$\rho_{f\,ij} = \frac{1}{(N-1)} \left[ abs\left( \sum_{k \neq i} \rho_{ik} \right) \right]^{\frac{1}{2}} \left[ abs\left( \sum_{l \neq j} \rho_{lj} \right) \right]^{\frac{1}{2}} \operatorname{sgn}\left[ \sum_{k \neq i} \rho_{ik} \right] \operatorname{sgn}\left[ \sum_{l \neq j} \rho_{lj} \right] \tag{9.5}$$

Visually, this amounts to approximating the original correlation matrix element $\rho_{ij}$ by a geometric average of signed averages over the matrix elements of the row and column whose intersection contains $\rho_{ij}$. Note that the approximation can have either positive or negative sign. Since $c_i$ is obtained using an average, the extent to which $\rho_f$ approximates $\rho$ will depend on the internal differences of matrix elements of $\rho$, with relatively constant matrix elements being more tractable.

A least-squares approach for getting the $\{c_i\}$ may also be useful. These equations are nonlinear, and need to be solved numerically.

Regardless, the requirement that $\rho_f$ is positive definite needs to be checked explicitly.

## 10. Applications to Options

The perturbation expansion has potential application to options that are dependent on several variables. For simplicity, we restrict the discussion to an illustrative example.

Consider the expectation of a "payoff" function $C(\boldsymbol{x})$ of the variables $\{x_i\}$ such that the limits $\{x_i^{\max}\}$ remain constants, and not interlocked[2], independent of the variables $\{x_i\}$. We can use either the Gaussian or Student-t multivariate distribution, depending on the correlation matrix $\rho$. Carrying out the same perturbation procedure as above (for which $C(\boldsymbol{x})=1$) leads to a perturbation series for this expectation of $C(\boldsymbol{x})$ using the $\rho_f$-dependent approximate measure.

Recall the general nature of the expectation $\langle F \rangle_{\rho_f}$ with respect to the $\rho_f$-dependent approximate measure in Eq. (5.6). We now want to use

$$F\left[\{x_i\}\right] = C(\boldsymbol{x}) \sum_{\beta=0}^{\infty} \frac{1}{\beta\,!} \left(-\tfrac{1}{2}\right)^{\beta} \left(\boldsymbol{x}^T \varepsilon \boldsymbol{x}\right)^{\beta} \tag{10.1}$$

We go through the same procedure as before. Similarly to Sect. 6 and 7, the functions for a given approximation at $\beta = 0,1,2$ etc. need to be evaluated including the dependence on the payoff function $C(\boldsymbol{x})$.

## 11. Application to Credit Risk and Credit Derivatives

The perturbation expansion developed in this paper can be potentially useful for analysis for credit risk and credit derivatives[vi].

For buy-and-hold portfolios, the determination of credit risk obtained using historical correlations between asset returns in a "structural model" can be appropriate. Analysis with these large correlation matrices could potentially be numerically handled using the perturbation methods in this paper.

For trading portfolios, "reduced-form models" that calibrate to market CDS, CDO prices and other data are appropriate. If the reduced-form model has several factors, we could then use the one-factor perturbation expansion to provide a potentially useful evaluation tool for credit derivatives.

[2] **Options:** This assumption is true for some, but not all, options. Some options with min/max conditions that interlock the limits through dependence on the $\{x_i\}$ can be handled by changing variables in the integrals, before applying perturbation theory.

## 12. Correlation Sensitivity

We get straightforward approximations for correlation sensitivity using the perturbation expansion that may prove useful. Consider a function $C(\rho)$ of the correlation matrix. Suppose the correlation matrix changes from $\rho = \rho^{(1)}$ to $\rho = \rho^{(2)}$. The resultant change in the function, i.e. the correlation sensitivity, is

$$\Delta_\rho C = C\left(\rho^{(2)}\right) - C\left(\rho^{(1)}\right) \tag{12.1}$$

Suppose further that we choose the approximate matrix $\rho_f^{(2)}$ for $\rho^{(2)}$ as unchanged from the approximate matrix $\rho_f^{(1)}$ of $\rho^{(1)}$, that is $\rho_f^{(2)} = \rho_f^{(1)} = \rho_f$. Then subtracting the two perturbation expansions we get an approximation to the correlation sensitivity. In first order $\beta = 1$, the $\rho_f$ dependence cancels out except for the approximate measure.

## 13. Convergence of the Perturbation Series

In this Section we discuss convergence of the perturbation series, although we have not constructed a formal proof. There are two main points. First, as a function of the order $\beta$, the exponential expansion contains a factor $(\beta!)^{-1}$, which eventually overwhelms powers $(Q)^\beta$ for any $Q$ as a function of $\beta$. Second, rapid decrease of the integrand at large values of $\varsigma$ exists, which means that the integrals are effectively finite, so the interchange of $\beta$-summation and $\varsigma$-integration is allowed.

Some intuition can be gained by looking at the 0th-order approximation. Define average parameters $c_{avg}$ and $x_{avg}^{\max}$ (the details will not matter). The $\varsigma$ dependence of the integrand is essentially of the form $\exp\left(\Phi(\varsigma)\right)$ where

$$\Phi(\varsigma) = -\tfrac{1}{2}\varsigma^2 + N \log\left[\mathfrak{N}\left(\frac{x_{avg}^{\max} - c_{avg}\varsigma}{s_{avg}}\right)\right] \tag{13.1}$$

Again, $N$ is the dimension and $\mathfrak{N}(\ )$ is the normal integral. We differentiate Eqn. (13.1) and set $\Phi'(\varsigma^*) = 0$ at the point $\varsigma = \varsigma^*$ where the magnitude of the integrand is largest. The equation to determine $\varsigma^*$ is

$$\varsigma^* = -\frac{Nc_{avg}}{s_{avg}} \mathfrak{N}'\left(\frac{x_{avg}^{\max} - c_{avg}\varsigma^*}{s_{avg}}\right)\left[\mathfrak{N}\left(\frac{x_{avg}^{\max} - c_{avg}\varsigma^*}{s_{avg}}\right)\right]^{-1} \tag{13.2}$$

Here, the derivative of the normal integral is $\mathfrak{N}'(\ )$. The 0th-order integrand has a pronounced bump at $\varsigma = \varsigma^*$, falling off rapidly on either side. Similar remarks hold for the 1st and 2nd order integrands, proportional to $\mathfrak{N}$ and $\mathfrak{N}'$, although due to cancellations between terms there may be some additional structure in $\varsigma$ besides a simple bump. As $N$ increases, it would appear that $\varsigma^*$ will increase too, but the other factors can decrease to make $\varsigma^*$ relatively constant in $N$.

Given the above bump structure, it is intriguing to speculate that a WKB approximation might be useful as a rough guide, without doing any integrations.

It is also instructive to look at the dominant order $\beta^*$ at fixed dimension $N$. Arguments similar to the above yield a rough estimate for $\beta^*$ as

$$\beta^* = \tfrac{1}{2} N^2 \varepsilon_{\text{avg}} \tag{13.3}$$

where $\varepsilon_{\text{avg}}$ is an average matrix element of the $\varepsilon$ matrix. Finally, at a given order $\beta$, the most important dimension $N^*$ is roughly given by

$$N^* = \frac{2\beta}{\ln\left(1/\mathfrak{N}_{\text{avg}}\right)} \tag{13.4}$$

where $\mathfrak{N}_{\text{avg}}$ is a normal integral at $\varsigma = \varsigma^*$ with average parameters as above.

## *Metrics for Convergence*

The theory in this paper is exact if all orders in perturbation theory are utilized. However, the whole idea is to stop at some manageable point, e.g. 2nd order. Hence there will be some error. We need to characterize the error to understand when the approximation can be useful. To this end, we define some metrics; the numerical error will be a function of these metrics.

Since we are using a one-factor approximation with a limited number of variables $\{c_i\}$, if the correlation matrix elements are approximately constant the approximation will be good, and the amount to which the correlation matrix elements are highly variable will limit the utility of the approximation. Hence, we anticipate that the internal correlation variance $\sigma^2_{\rho,\text{Int}}$ of $\rho$ will be a useful metric, where

$$\sigma^2_{\rho,\text{Int}} = \left[\tfrac{1}{2} N(N-1) - 1\right]^{-1} \sum_{i<j}^{N} \left(\rho_{ij} - \rho_{\text{avg}}\right)^2 \qquad (13.5)$$

Here, $\rho_{\text{avg}}$ is the average of the off-diagonal matrix elements of $\rho$. In particular, since the approximate matrix $\rho_f$ is obtained through averaging over elements of $\rho$, its internal variance $\sigma^2_{\rho_f,\text{Int}}$ will be smaller than $\sigma^2_{\rho,\text{Int}}$. We also anticipate that $\sigma^2_{\varepsilon,\text{Int}}$ , the internal variance of the expansion matrix $\varepsilon = \rho^{-1} - \rho_f^{-1}$, will be useful.

If $\rho$ is close to being singular (i.e. has one or more very small eigenvalues), then $\rho^{-1}$ will have large matrix elements, but $\rho_f^{-1}$ will not. Therefore $\varepsilon = \rho^{-1} - \rho_f^{-1}$ will have large elements and convergence will be less rapid. Cutoffs on the minimum eigenvalue may have to be applied in some cases,

$$\lambda^{(\mu)} > \lambda^{(\text{Min})}(N) \qquad (13.6)$$

That is, the correlation matrix may need to be regularized somewhat in order to apply this method. Naturally, the correlations in the regularized matrix will be changed somewhat, which is a drawback. On the other hand, it is known that correlations in practice are highly unstable with large uncertainties[iii], so the change due to regularization may be small compared to these uncertainties. In any case, this procedure must be consistent with the experimental uncertainties of the correlation matrix elements. We anticipate that the cutoff will depend on the dimension $N$ .

A useful metric for the correlation matrix is the distance of the matrix from a singular matrix, which we call $R(N)$. We suggest $R(N)$ be defined as

$$R(N) = \frac{1}{\sum_{\mu=1}^{N} \left( \frac{1}{\lambda^{(\mu)}} \right)} \qquad (13.7)$$

Eqn. (13.7) is formally the same as the total resistance of $N$ resistors in parallel, with the resistances taken as $\left\{\lambda^{(\mu)}\right\}$. If some $\lambda^{(\mu)} \to 0$, then $R(N) \to 0$ and the matrix becomes singular. For example, if all correlation matrix elements are equal to one, then all but one eigenvalue are equal to zero. A zero eigenvalue in the resistor circuit analogy corresponds to a "short circuit".

The maximum distance of the matrix from the singular case occurs when all eigenvalues are equal. Since we must have $\sum_{\mu=1}^{N} \lambda^{(\mu)} = N$, we get $\lambda^{(\mu)} = 1$, and so $\text{Max}\left[R(N)\right] = \frac{1}{N}$. Note that the distance from a singular matrix decreases as the dimension increases.

## 14. Enhancing the Convergence with Padé Approximants

In this Section we discuss enhancing the convergence of the perturbation series using Padé approximants. This is a common procedure in scientific applications, but it may not be so familiar in finance, and so we shall spend a little time giving some background. We shall also use the idea in a somewhat extended fashion.

A Padé approximant is defined as a rational function that agrees with the perturbation expansion of a function to a given order. Padé approximants[vii] can enhance numerical convergence. In some cases (including applications of this theory), Padé approximants may be more than numerically helpful; they may be necessary.

For small values of matrix elements $\varepsilon_{ij}$ of $\varepsilon = \rho^{-1} - \rho_f^{-1}$ in which the perturbation is being carried out, the perturbation expansion will converge rapidly, and no problems are anticipated. However, for larger values of $\varepsilon_{ij}$, the convergence will be less rapid, or even appear to diverge at some finite order. We have calculated through 2$^{nd}$ order. We claim that the perturbation series converges. Even if the perturbation series were to diverge, for example as an asymptotic series, the Padé approximants provide a method of summing the series.

Consider the simple case with equal matrix elements $\varepsilon_0$ (i.e. $\varepsilon_{ij} = \varepsilon_0$ for $i \neq j$), so the perturbation expansion is in $\varepsilon_0$. Simplifying the notation, we have $I^{(\beta)} = O\left(\varepsilon_0{}^{\beta}\right)$ for the $\beta$ th order perturbative term with $\beta = 0,1,2$. The series expansion of the Padé approximant $I_{\text{Pade}}^{(\beta)}$ matches the sum of the perturbative terms to order $\beta$, viz $I^{(\beta)}_{\text{Pade}} = \sum_{\gamma=0}^{\beta} I^{(\gamma)} + O\left({\varepsilon_0}^{\beta+1}\right)$. As $\beta$ increases, the match to the perturbation series becomes more exact. The expansion can be carried out at small values of $\varepsilon_0$, and then the Padé approximants can be analytically continued to large values of $\varepsilon_0$. The Padé approximants effectively provide approximations to arbitrarily high-order perturbative terms without explicitly calculating them. Again, the Padé results can be numerically better than the perturbation expansion. Padé approximants are labeled by $(L, M)$ or $[L/M]$, where $L$ and $M$ are the orders of the numerator and denominator.

So, we define successive Padé approximants $I^{(\beta)}_{\text{Pade}}$ of 0$^{\text{th}}$, 1$^{\text{st}}$, and 2$^{\text{nd}}$ order as:

$$I^{(0)}_{\text{Pade}} = I^{(0)} \tag{14.1}$$

$$I^{(1)}_{\text{Pade}} = I^{(0)}\left[1 - \frac{I^{(1)}}{I^{(0)}}\right]^{-1} \tag{14.2}$$

$$I^{(2)}_{\text{Pade (1,1)}} = \left[I^{(0)} + I^{(1)} - \frac{I^{(0)}I^{(2)}}{I^{(1)}}\right]\left[1 - \frac{I^{(2)}}{I^{(1)}}\right]^{-1} \tag{14.3}$$

The above 2$^{\text{nd}}$-order Padé approximant is of $[1/1]$ type. Another 2$^{\text{nd}}$-order approximant, of $[0/2]$ type, is defined as

$$I^{(2)}_{\text{Pade (0,2)}} = I^{(0)}\left[1 - \frac{I^{(1)}}{I^{(0)}} - \frac{I^{(2)}}{I^{(1)}} + \left(\frac{I^{(1)}}{I^{(0)}}\right)^2\right]^{-1} \tag{14.4}$$

Often the diagonal approximants $[L/L]$ are used, but off-diagonal approximants can be useful. Note that $I^{(1)}_{\text{Pade}}$ has type $[0/1]$.

We will use these same Padé approximant formulae for the real case of non-constant $\varepsilon_{ij}$ matrix elements, dropping the restriction of constant $\varepsilon_0$. This procedure goes beyond the usual Padé theory. However, the assumption is simple to write down, and it can be tested numerically against direct integration.

An additional acceleration device may turn out to be useful. We can try to parametrically and approximately extrapolate the Padé approximants to $\beta \to \infty$, i.e. infinite order in $\beta$, thereby obtaining $I^{(\infty)}$ for the final approximation to the original multivariate Gaussian integral, Eqn. (2.1). Consider the ansatz

$$I_{\text{Pade}}^{(\beta)} = \left[ I^{(0)} - I^{(\infty)} \right] e^{-\alpha\beta} g(\beta) + I^{(\infty)} \qquad (14.5)$$

Here $\alpha$ is a parameter and $g(\beta)$ is some reasonable function, with $g(0)=1$ so that $I_{\text{Pade}}^{(0)} = I^{(0)}$. We might choose $g(\beta)=\cos(\pi\beta)$ [or $g(\beta)=1$] for an oscillating [non-oscillating] Padé sequence, which is known at $\beta = 0,1,2$. We know $I_{\text{Pade}}^{(1)}$ and $I_{\text{Pade}}^{(2)}$ (see footnote[3]). Hence there are 2 equations and 2 unknowns, and hence we can get the final result $I^{(\infty)}$ by determining $\alpha$.

## 15. Numerical Aspects, Logical Flow Chart

Numerical aspects of the theory presented in this paper are currently being investigated, including error analysis as a function of parameters, and will be reported separately [viii]. Preliminary results are encouraging, but also indicate that care is required in applying the theory, including Padé approximants.

Here is a logical "flow chart" with explicit steps in order to obtain the Gaussian $I_N\left[\{x_i^{\max}\}; \rho_f\right]$ in Eqn. (2.1), using the 1-factor approximation.

1. Obtain the $N$-dimensional correlation matrix $\rho$. Check that $\rho$ is positive definite (PD). It may be necessary to impose an eigenvalue cutoff $\{\lambda^{(\mu)} > \lambda^{(\text{Min})}(N)\}$ to make the 2nd-order approximation "accurate enough". Get the inverse $\rho^{-1}$ and determinant $\Delta(\rho)$.
2. Get $\{c_i\}$ via Eq. (9.4), or least squares, from $\rho$; write $s_i = +\sqrt{1-c_i^2}$.
3. Get the one-factor approximation $\rho_f$ to $\rho$ in Eqn. (9.5). Check that $\rho_f$ is PD. Calculate the inverse matrix $\rho_f^{-1}$ in Eqns. (5.10) - (5.11), the determinant $\Delta(\rho_f)$, and $J(\rho_f, \rho) = \sqrt{\Delta(\rho_f)} \big/ \sqrt{\Delta(\rho)}$.
4. Get $\varepsilon = \rho^{-1} - \rho_f^{-1}$ (subtract the matrices element by element).
5. Specify the upper limit values $\{x_i^{\max}\}$.

[3] **Which 2nd-order Padé?** It may be useful to take the maximum or the average of the two 2nd-order Padé approximants.

6. For each $\varsigma$ in the 1-factor integration $\varsigma \in (-\Lambda, +\Lambda)$ where $\Lambda$ is a "sufficiently big number", calculate for each $i = 1...N$ :

    - $\xi_i^{\max}(\varsigma)$ in Eqn. (2.5)
    - $\nu_i(\varsigma) = \mathfrak{N}\left[\xi_i^{\max}(\varsigma)\right]$ using a normal integral approximation[ii]
    - $\chi_i(\varsigma),\ w_i^{(1)}(\varsigma),\ w_i^{(2)}(\varsigma)$ in Eqns. (6.2) - (6.4)
    - $G_{ij}(\varsigma)$ in Eqns. (6.5) - (6.6) for the 1st order contribution $I_N^{(\beta=1)}\left[\{x_i^{\max}\}; \rho_f; \rho\right]$
    - $w_i^{(3)}(\varsigma),\ w_i^{(4)}(\varsigma)$ in Eqns. (7.1) - (7.2)
    - $G^{(\Gamma)}_{(ijkl) \text{ in } C(\Gamma)}$ in Eqns. (7.1) - (7.9) for the 2nd order contribution $I_N^{(\beta=2)}\left[\{x_i^{\max}\}; \rho_f; \rho\right]$

7. Calculate 0th order contribution $I_N^{(\beta=0)}$ in Eqn. (2.10).
8. Calculate 1st order contribution $I_N^{(\beta=1)}$ in Eqn. (2.11) and obtain the 1st-order approximation $I_N^{(\text{1st Order Approx.})} = I_N^{(\beta=0)} + I_N^{(\beta=1)}$ .
9. Calculate 2nd order contribution $I_N^{(\beta=2)}$ in Eqn. (2.12) and obtain the 2nd-order approximation $I_N^{(2nd\ \text{Order Approx.})} = I_N^{(\beta=0)} + I_N^{(\beta=1)} + I_N^{(\beta=2)}$ .
10. Calculate the $\beta = 1, 2$ Padé approximants $I_{\text{Pade}}^{(\beta)}$ in Eqns. (14.2) - (14.4).
11. Obtain the extrapolation of the Padé approximants, $I^{(\infty)}$, as $\beta \to \infty$ using Eqn. (14.5). $I^{(\infty)}$ is the suggested final result.
12. Use the internal correlation variance $\sigma^2_{\rho,\text{Int}}$ and the distance from a singular matrix $R(N)$ from Sect. 13 as metrics for application analysis.

## Acknowledgements

I thank Kris Kumar for discussions regarding numerical aspects and potential applications of the theory.

File =
DASH Multivariate Integral Perturbative Techniques I Theory Sept06 R2 posted.doc
Last accessed: 10/13/2006, 7:46 AM